
\font\twelvebf=cmbx12
\font\ninerm=cmr9
\nopagenumbers
\magnification =\magstep 1
\overfullrule=0pt
\baselineskip=18pt
\line{\hfil CCNY-HEP 5/94}
\line{\hfil May 1994}
\vskip .8in
\centerline{\twelvebf Wave Functionals, Hamiltonians and the }
\centerline{\twelvebf Renormalization Group}
\vskip .5in
\centerline{\ninerm D. MINIC and V.P. NAIR}
\vskip .1in
\centerline{ Physics Department}
\centerline{City College of the City University of New York}
\centerline{New York, New York 10031.}
\centerline {E-mail: minic@scisun.sci.ccny.cuny.edu}
\centerline{~~~~~~~~~~~vpn@ajanta.sci.ccny.cuny.edu }
\vskip 1in
\baselineskip=16pt
\centerline{\bf Abstract}
\vskip .1in
We analyze the renormalization of wave functionals and energy eigenvalues
in field theory.
A discussion of the structure of the renormalization group equation
for a general
Hamiltonian system is also given.
\vfill\eject

\footline={\hss\tenrm\folio\hss}

\def \12 {{\textstyle {1\over 2}}}
\def \L {{\Lambda}}

\magnification =\magstep 1
\overfullrule=0pt
\baselineskip=22pt
\pageno=2
\noindent{\bf 1. Introduction}
\vskip .2in
The Schr\"odinger representation of a field theory, using wave functions
which are functionals of the field variables, has been known for a long
time [1]. Nevertheless, it
has rarely been used since other
representations are far more convenient
for most field theoretic calculations. However, there are many problems,
especially those involving changes in ground state properties, for
which our intuitive understanding is better served in a Hamiltonian
framework and using wave functionals. Analysis of Quantum Chromodynamics
in lightcone framework [2], saturation problems in deep inelastic scattering
[3]
and many problems in condensed matter physics may belong to this category.
Wave functionals may also be useful in a variational approach.
The most obvious difficulty in using wave functionals in field theory is in
understanding their renormalization properties, especially  in the modern
or Wilsonian interpretation, where we consider how various quantities must
change as high momentum modes are progressively integrated out [4]. Actually,
some time ago, motivated by application to the Casimir effect, Symanzik
analyzed a closely related question [5]. He considered the Euclidean partition
function in a space with a boundary and used combinations of Dirichlet and
Neumann-type Green's functions to ensure proper behaviour of field variables
on the boundary. Renormalization can be carried out by introducing
counterterms which are monomials of the field and their derivatives on the
boundary, in addition to the usual counterterms in the action.
Although the results can be reinterpreted using wave
functions [5,6], the analysis is not in the context of partially
integrating out high
momentum modes. In this paper, we study the change of wave functionals
as high momentum modes are integrated out and obtain the corresponding
renormalization group (RG) equation. This may also be the first step in
alleviating some of the difficulties with variational wave functions
in field theory [7].

In the next section we shall consider the theory of a real scalar field
since this suffices to illustrate most of the general ideas. In section 3,
we show explicitly the computation
of the renormalization constants related to the surface counterterms.
The RG-equations for wave functionals and energy eigenvalues are obtained
in section 4. Although these equations are derived for our
example of a scalar field, the structure is quite general
and will hold in any field theory. The last section gives a discussion of the
RG-equation for a general Hamiltonian system.
\vskip .2in
\noindent{ \bf 2. The reduced wave functional}
\vskip .2in
Consider the theory of a real scalar field $\phi (x,t)$
with $Z_{2}$ symmetry with an action of the form
$$
S(\Lambda, \phi ,t,t_0 )= \int_{t_0}^t  dx^0\int d^3x~ Z_3(\L )
\left [ \12 {(\partial \phi )^2}~+ \12
(m^2 -\delta m^2)\phi^2 \right] ~+ Z_1(\L )\lambda \phi^4 \eqno(1)
$$
$\Lambda$ denotes the upper cutoff on the momenta.
The wave functions of the theory are functionals of the
field $\phi$ at the fixed time $t$. We decompose the original field into
high and low momentum modes as $\phi(x,t)=\chi(x,t) + \varphi(x,t)$, where
$\chi$ denotes modes of high momenta (heavy modes), i.e. with momenta
constrained to a shell $\mu <k \leq \Lambda$,
and $\varphi$ denotes modes of low momenta (light modes) with $k\leq \mu$.

Integration over the heavy modes
generates an effective field theory with a cutoff $\mu$. Of course,
this integration entails also the redefinitions of parameters so that
the low energy correlation functions are preserved, i.e., they are
invariants of the renormalization group (RG) flow.
In terms of the wave function, this must produce a low energy wave function
$\Psi_{red} (\varphi )$ starting from $\Psi (\phi)= \Psi[ \chi , \varphi ]$.
As we shall see below in explicit calculations, this reduced wave function
has the form
$$
\Psi_{red} (\mu, \varphi, t ) ~= U[\mu, \varphi ]\int [d\varphi ]~
 e^{iS (\mu, \varphi , t, t_0 )}
\Psi_{red}(\mu, \varphi, t_0) \eqno(2)
$$
where $U$ is given by
$$
U[\mu, \varphi ] ~= \exp  i \int d^3x~ \left[ Z_4(\mu ) \varphi \partial_t
\varphi
{}~+ \delta (\mu)\varphi^2 \right]~\equiv e^{iG (\mu )}\eqno(3)
$$
$U$ compensates for possible cutoff dependent surface terms ( on
the time-slice at $t$) generated by the integrations over $\chi$-modes.
$S (\mu, \varphi)$ is given by Eq.(1) with a cutoff $\mu$ instead of
$\L$. For the wave function,
we need new
renormalization constants $Z_4$ and $\delta$. The integration in Eq.(2) is over
all values of the fields between $t_0$ and $t$, and over the fields at
$t_0$, as usual.

Now, integration over the heavy modes, at least diagrammatically, means that we
allow $\chi$-fields only on internal lines or propagators. Such propagators
represent vacuum fluctuations of the $\chi$-field. Hence we can obtain the
reduced wave function by treating the $\chi$-modes as a kind of environment
to which the $\varphi$-modes are coupled, the distribution of the
heavy modes in the environment being characterized by
a vacuum wave function of the $\chi$-modes, say $\Psi_{0}[\chi]$.
Thus we start with $\Psi (\L )$ which has a representation as in Eq.(2), i.e.,
$$
\Psi (\L, \phi )=
U[\L, \phi ]\int [d\phi ]~
 e^{iS (\L, \phi , t, t_0 )}
\Psi(\L, \phi, t_0) \eqno(4)
$$
Writing $\phi =\chi + \varphi$, the action splits as
$$
S(\phi)=S(\chi)+ S(\varphi )~+ S_{int}(\chi ,\varphi). \eqno(5)
$$
We define a function
$$
F(\chi ,\varphi ,t ) \equiv \int [ d\chi ]~\exp i\left[ S(\chi ,t,t_0 )+
S_{int}(\chi ,\varphi ,t,t_0) \right] ~\Psi (\chi, \varphi, t_0)\eqno(6)
$$
The initial wave function can be assumed to factorize into a $\chi$-dependent
term and a $\varphi$-dependent term, at least as $t_0 \rightarrow -\infty$;
this
is essentially the Born-Oppenheimer approximation.
$\Psi_0[\chi]\equiv U[\chi, \varphi=0]~ F(\chi , \varphi=0 )$ then defines the
``vacuum wave
function" for the $\chi$-modes. The integration over $\chi$-modes can then be
represented on the wave function by $\Psi(\phi , \L)\rightarrow \Psi_{red}
(\varphi ,\mu)$ with
$$
\Psi_{red}(\varphi, \mu)~= \int [d\chi]~ \Psi_0^*(\chi,\L, t ) ~\Psi(\chi,
\varphi, \L, t) ~~\equiv ~<\Omega_\chi \vert \chi, \varphi >\eqno(7)
$$
This formula defines the procedure of integrating out the high momentum
modes and, along with (2), will lead to the computation of the renormalization
constants $Z_4$ and $\delta$.

Consider now the time-evolution of $\Psi_{red}$ as given by Eq.(2). The
quantity
$U[\varphi ]$ involves variables on the time-slice at $t$ and can be
interpreted in the operator language as a unitary transformation.
Keeping this in mind, the variation of the
time-argument in Eq.(2) gives
$$
i \partial_{t} \Psi_{red}(\varphi,\mu ,t)=H_{red}(\varphi,\mu)
\Psi_{red}(\varphi,\mu ,t) \eqno(8)
$$
where
$$
H_{red} = U~H(\varphi )~U^{-1} \eqno(9)
$$
and $H(\varphi )$ is the Hamiltonian operator given by the action
$S(\mu, \varphi, t,t_0)$; $H(\varphi)$ includes some of the effects of
integrating
out the $\chi$'s, viz. those associated with spacetime volume integrals.
By rewriting $\Psi_{red}$ using Eq.(7), we see that
$$
i \partial_{t} \Psi_{red}(\varphi,\mu)=<\Omega_{\chi}|-H_{\chi}|\chi,\varphi>+
<\Omega_{\chi}|H(\chi,\varphi)|\chi,\varphi> \eqno(10)
$$
If the $\chi$-modes have momenta very high compared to the $\varphi$'s
of interest, they cannot be excited by the $\varphi$'s for kinematical
reasons. We can expand
$\vert \chi, \varphi >$ in terms of a complete set of $\chi$-states and
we then see that
the above matrix elements vanish except for the case $ \vert \chi,\varphi >
= \Psi_0(\chi) \vert \varphi>$, i.e., $\chi$-modes have only
vacuum fluctuations. Eq.(10) thus gives $H_{red}$ as
$$
H_{red}(\varphi,\mu)=<\Omega_{\chi}|H(\chi,\varphi)|\Omega_{\chi}>-
E_{\Omega_{\chi}} \eqno(11)
$$
If $\chi$'s are thought of as
environmental degrees of freedom, we expect the effective
Hamiltonian for $\varphi $'s to be $H(\chi ,\varphi )$ averaged over the
distribution of $\chi $'s, exactly as in Eq.(11). We now see that such a
Hamiltonian does indeed generate
the time-evolution of the reduced wave function as defined in Eq.(2).
\vskip .2in
\noindent {\bf 3. Explicit calculation of the reduced wave function}
\vskip .2in
We now proceed to calculate explicitly the reduced
wave function as defined by Eq.(7). There will be the usual corrections to the
action which
depend on the cutoff and which are spacetime volume integrals. Here we
are primarily interested in the surface terms which contribute to
$U[\varphi]$ and hence to $ Z_4,~\delta $. We shall calculate these to first
order
in $\lambda$.

The $\varphi$-modes play the role
of background fields as far as the calculation of $ F(\chi, \varphi ) $
is concerned. From its definition, we can write down an equation for
$ F(\chi, \varphi )$ as
$$
i \partial_{t} F =  \left[ ~\int \Bigl( \12 Z_3^{-1} \pi_{\chi}^{2}+ \12
 Z_3 (\nabla \chi)^{2}+ \12 Z_3 (m^2-\delta m^2)\chi^2 \Bigr)
 ~+H_{int}(\chi,\varphi) + H_{int}(\chi)\right] ~~F
\eqno(12)
$$
For calculations to first order in $\lambda $, we can simplify this as
$$
i \partial_{t}F \approx \12 \int d^3x~ [-{{\delta^2}\over{\delta \chi^2}} +
\omega^2 \chi^2 + V \chi^2]~F
 \eqno(13)
$$
where $\omega^2 \chi^2 = \chi (-\nabla^2 +m^2) \chi$ and
$V=12\lambda \varphi^2$. $V$ varies slowly in time compared to $\chi$.
Thus an adiabatic approximation is suitable and we can use a WKB-type ansatz
$$
F= \exp (i(S^{(0)}+S^{(1)}+...)) \eqno(14)
$$
with
$\partial_{t} S^{(0)} \approx {\cal O}(1),~\delta_{\chi} S^{(1)}
\approx {\cal O}(1)$ , etc.
To the order we are interested in, we get
$$
F\approx \exp \left[-\12 \int \chi (\sqrt{\omega^2 +V})\chi  -
{\textstyle{{i}\over{2}}}\int_{t_0}^t dt~
\int \sqrt{\omega^2 +V} ~+{\textstyle{i \over 4}}\int \chi
{{1}\over {\omega^2 +V} }{\partial_t V}\chi ~+...\right] \eqno(15)
$$
The reduced wave function has the form
$$
\Psi_{red}=\int [d \varphi] \exp(iS(\varphi)) \int F^{*} (\chi, \varphi=0)
 F(\chi, \varphi)d\chi \eqno(16)
$$
By expanding $F(\chi ,\varphi)$ in a series in $V$ we obtain
 a linearly divergent surface term
$$
\12 <\chi {V \over 2\omega}\chi>={3\lambda \over 4\pi^2}(\Lambda-\mu)
\int \varphi^2 d^3x \eqno(17a)
$$
and a  logarithmically divergent surface term
$$
{i \over 8}<\int \chi {{\dot V}\over {\omega^2}} \chi>~=i {3\lambda \over
8\pi^2}
 \log({\Lambda^2}/{\mu^2})\int \varphi \partial_t \varphi d^3 x  ,\eqno(17b)
$$
where $<X>= \int d\chi~F^*(\chi,\varphi=0)XF(\chi, \varphi=0)$.
Collecting terms we get to
first order in $\lambda$
$$
iG(\mu) = iG(\L )~+{3 \lambda \over 4\pi^2}(\Lambda-\mu)  \int d^3x~ \varphi^2~
 +
i{{3\lambda}\over {8\pi^2}}\log({\Lambda^2}/{\mu^2})
 \int d^3x~ \varphi \partial_t \varphi +...\eqno(18)
$$
For the $\L$-dependence to cancel out we need
$$\eqalign{
Z_4 (\L) ~+ {3\lambda \over {8\pi^2}} \log(\L^2 /\mu^2)~&= Z_4 (\mu)\cr
i\delta (\L) ~+ {3\lambda \over 4\pi^2}(\L -\mu)~&=i\delta (\mu)\cr}\eqno(19)
$$
or in other words,
$$\eqalign{
Z_4 (\L )~&= c_4 ~- {3\lambda \over 8\pi^2} \log \L^2 ~+ ... \cr
i\delta (\L)~& = c_0 ~-{3\lambda \over 4\pi^2} \L ~+...\cr}
\eqno(20)
$$
The constants $c_4,~c_0$ depend on the normalization conditions chosen;
a minimal choice is to set them to zero. (Symanzik used $c_4 =1$.)

One can give general power counting arguments, along the lines of ref. [4], to
see that the surface counterterms we have introduced suffice to absorb
all the $\L$-dependence.
\vskip .2in
\noindent{\bf 4. RG Equation for $\Psi$}
\vskip .2in
The wave function (3) is independent of $\L$, as seen from the explicit
computations, or equivalently, $\Psi$ in Eq.(2) is independent of $\mu$ when
further
integrations are carried out. This leads to the renormalization group equation.
In what follows, we shall be using reduced wave functions and Hamiltonians;
we shall leave this qualification as understood, omitting the subscript
$red$. Also we ignore the mass terms for simplicity; they can be restored
in a straightforward fashion if desired.
Note that $Z_3\partial_t \varphi$ is the canonical momentum $\pi$. We may
thus write
$$
G ~= iZ_4Z_3^{-1}\int d^3x~\varphi ~\pi \eqno(21)
$$
Consider an infinitesimal change in the cutoff $\L$ in the action, say  $\L
\rightarrow \L (1-\epsilon)$. After Fourier transformation and rescaling of
the corresponding momenta, we find
$$
-{dS \over {d\log{\Lambda}}}=\int d^4x~(x\cdot \nabla \varphi + \varphi +\gamma
\varphi ) {\delta \over \delta \varphi}S-\beta_{i}
{\partial \over \partial \lambda_{i}}S
+\int Hdt \eqno(22)
$$
The origin of the last term ($\int Hdt$) is due to the fact that
the integration over time is only upto $t$.
In Eq.(22), we have introduced both the $\beta$-function and the anomalous
dimension
$\gamma$, defined by
$$\eqalign{
\gamma &= -{1\over 2} \left[ {{\partial \log{Z_{3}}}\over
{\partial \log{\Lambda}}}\right]_{\lambda^{(0)}}\cr
\beta & =-\left[ {{\partial \lambda_{i}}\over {\partial \log{\Lambda}}}
\right]_{\lambda^{(0)}} \cr}\eqno(23)
$$
where $\lambda^{(0)}$ is the bare coupling constant, defined by $Z_1 \lambda=
Z_3^2 \lambda^{(0)}$.

Let us now consider the surface term $Z_{4}Z_{3}^{-1}\int \varphi \pi$. It is
easy to see, again by Fourier transformation followed by rescaling of momenta,
that $\int \varphi~\pi$ is independent of $\L$; the contribution to the
variation
of the wave function from $G(\L)$ is thus solely due to $Z_4Z_3^{-1}$.
In using Eq.(22) we need the partial integration formula
$$
\int [d\varphi ]~\left[ \int i(x\cdot \nabla \varphi +\varphi +\gamma \varphi)
{\delta S \over \delta \varphi}\right] \exp(iS)=\left[
\int d^3x~(x\cdot \nabla \varphi +\varphi +
\gamma \varphi ){\delta \over \delta \varphi}\right]~(\int [d\varphi ]
\exp(iS)) \eqno(24)
$$
On the right hand side, the field variables and functional derivative refer to
the values on the boundary, i.e., at time $t$. Using this equation and the
$\L$-independence of $\Psi$, we immediately get the renormalization group
equation
$$
-{{d\Psi}\over {d\log \L}}=
\left[ ~i\int d^3x~(x\cdot \nabla \varphi +(1+\gamma +\sigma) \varphi)]~\pi ~-
 \beta {\partial \over \partial \lambda} +i H~t \right]~\Psi =0.
\eqno(25a)
$$
or equivalently
$$
\left[ ~i \int d^3x~(x\cdot \nabla \varphi +(1+\gamma +\sigma) \varphi) \pi
-\beta
{\partial \over \partial \lambda} -t {\partial \over \partial t}\right]~
 \Psi =0 \eqno(25b)
$$
where
$$
\sigma \equiv -\left[ {{\partial (Z_4 Z_3^{-1})}\over {\partial
\log \L}}\right]_{\lambda^{(0)}}~={ {3\lambda}\over {4\pi^2}}~+...\eqno(26)
$$
is the new anomalous dimension arising due to the logarithmically divergent
surface term. This result agrees with Symanzik [4].

Strictly speaking, our discussion so far applies to the vacuum wave function.
This is because we have neglected the nontrivial $\varphi$-dependence
and associated properties of the initial wave function in the
functional integral formulae. When we discuss excited states, we must have
a nonzero number of ``particles" in the initial state or they have to
be created at some time prior to $t$ which corresponds to the
insertion of field variables in the
functional integral. Either way, the partial integration formulae are
more subtle. Notice that it is $U^{-1} \Psi$ which is evolved by $e^{iS}$.
Thus Eq.(2) should more accurately read as
$$
\Psi (t) ~=~U(t) \int [d\varphi ]~ e^{iS(t,t_0)} \left( U^{-1}(t_0)
\Psi (t_0)\right)  \eqno(27)
$$
The cutoff-independence of the wave functions then leads to the
equation
$$
({\cal D} \Psi) (t) - U (t) \int e^{iS}~ U^{-1} (t_0)
({\cal D} \Psi) (t_0)=0 \eqno(28)
$$
where we have defined
$$
{\cal D} = ~ i \int d^3x~[x\cdot \nabla \varphi +(1+\gamma +\sigma)
\varphi]\pi-
 \beta {\partial \over \partial \lambda} +iHt \eqno(29)
$$
The question is whether we can set ${\cal D}\Psi$ to zero. Let us say that
$Q_\alpha$ are conserved quantities in the theory which can be used to label
the states, i.e., $Q_\alpha \Psi_q = q_\alpha \Psi_q$. Applying ${\cal D}$
on this equation, it is easy to see that ${\cal D}\Psi =0$ is inconsistent
unless $[ {\cal D}, Q_\alpha ]=0$. For the spatial momentum $P_i$, which
is conserved in the present case, we see that $[{\cal D}, P_i ]= P_i$ and we
cannot set $D\Psi$ to zero. From the
conservation property, we have
$$
p_{i}{{\partial \Psi_{p} \over \partial p_{i}}}=\exp(-iHt)p_{i}
{{\partial \Psi_{p}(t_{0})}\over {\partial p_{i}}}=U(t)~\int
e^{iS}~U(t_0)^{-1}p_{i}
{{\partial \Psi_{p}(t_{0})}\over {\partial p_{i}}} \eqno(30)
$$
Combining this with Eq.(28), we have
$$
({\cal D} \Psi +c~ p {{\partial \Psi}\over {\partial p}})= U\int
e^{iS}~U^{-1}(t_0)
 ({\cal D} \Psi +c ~p {{\partial \Psi}\over {\partial p}})(t_0)\eqno(31)
$$
It is consistent to set $({\cal D}+c~p{\partial \over \partial p})\Psi$ to zero
if we choose $c=1$. More generally, the RG-equation reads
$$\eqalignno{
\left[ {\cal D} ~+~ \sum_\alpha \left( h_\alpha q_\alpha
{\partial \over \partial q_\alpha }\right)\right] \Psi_q ~&=0&(32a)\cr
[ {\cal D}, Q_\alpha ] &=h_\alpha Q_\alpha &(32b)\cr}
$$
Specifically for the case of $N$ particles of momenta $p^{(m)},~m=1,2,...,N$,
we have
$$
\left[ \{ i\int d^3x~(x\cdot \nabla \varphi +  (1+\gamma +\sigma)\varphi)\pi
\} -
\beta {{\partial}\over {\partial \lambda}}-t {{\partial}\over {\partial t}} +
\sum_m~ p^{(m)}_{i} {{\partial}\over {\partial p^{(m)}_{i}}}\right]\Psi=0
\eqno(33)
$$
This gives the RG-flow of the wave function. For the special case of energy
eigenstates, viz., for
$$
\Psi=e^{-i\omega t} ~\psi [\varphi], \eqno(34)
$$
Eq.(33) also gives the flow of energy eigenvalue $\omega$ as
$$
\left[ \sum p \cdot {{\partial}\over {\partial p}}  -\beta {{\partial }\over
{\partial \lambda}}\right] \omega ~-\omega ~=0 \eqno(35)
$$
The solution to this equation is given, for example for a one-particle state,
in terms of an arbitrary function $f(s)$
as
$$
\omega={\sqrt{p^2}} ~f ( u+\log (p / \mu)) \eqno(36)
$$
where $du= {{d\lambda}\over {\beta(\lambda)}}$.
A simple application of this result is to the one-electron state in the
BCS theory of superconductivity
for which we have $\beta = -a \lambda^2 $ for the four-fermion coupling, with
$a=mk_{F}/{2\pi^2}$ [8]. ( Positive $a$ corresponds to asymptotic freedom. $m$
is the electron mass and $k_F$ is the Fermi momentum.)
In this case, $f= \log (p/\mu ) +{1/a\lambda}$, and if one assumes a mass gap,
$f(s)\approx e^{-s} $ for small $s$, so as to cancel the $p$-dependence,
eq. (36) then leads to the well-known relation
$ \omega=\mu \exp \left({-2\pi^2/{m \lambda k_{F}}}\right)$.

Our derivation of the RG-equations has been in the context of the scalar
field theory; nevertheless, it is clear that the general structure of
Eqs.(25), (33) and (35) will be obtained in any field theory.

\vskip .2in
\noindent{\bf 5. RG-equation  in Classical and Quantum Dynamics}
\vskip .2in
The basic idea of the renormalization group, viz. progressively integrating out
the high momentum modes is very general. It is of some interest therefore to
bring out the structure of the RG-equations already in classical dynamics.
Consider a Hamiltonian system with phase space variables $( \xi^i$ ,$\chi^i)$,
where $\chi$'s represent heavy modes or variables weakly coupled to
the ``light modes" $\xi^i$. One can integrate out the heavy modes to obtain an
effective Hamiltonian for the variables $\xi^i$. (A specific, purely classical,
context in which this can be concretely realized is in statistical mechanics
where
a Liouville distribution of $\chi$'s  can be given  and they can be integrated
out.) The change of the effective Hamiltonian as we lower the cutoff $\L$ is
what we are interested in. We write
$$
H[\Lambda(1-\epsilon)]=H(\Lambda)-\epsilon {{dH}\over {d\log{\Lambda}}}~
\equiv ~H+\epsilon V \eqno(37)
$$
The equations of motion with this Hamiltonian are
$$
{{d\xi^{i}}\over {dt}}=(\omega^{-1})^{ij}
{{\partial (H+\epsilon V)}\over {\partial \xi^{j}}} \eqno(38)
$$
where $\12 \omega_{ij}d\xi^{i}\wedge d\xi^{j}=\omega$ is the symplectic
two-form. In quantum field theory, we seek to preserve the correlation
functions
as we lower the cutoff. In the present case, what is preserved is the set of
solutions to the equations of motion. Thus one should be able to absorb the
change due to $V$ as a change of variables
$\xi^i \rightarrow \xi'^i = \xi^i +\epsilon v^i(\xi)$,
a change of parameters $g^\alpha  \rightarrow g'^\alpha = g^\alpha +
\epsilon \beta^\alpha $ and a change of scale for the time variable
$t \rightarrow t' =t (1- \epsilon h_H )$. ($h_H$ should be 1
for relativistic theories
with identical scaling of space and time variables. It can be different from
1 for nonrelativistic theories where the time variable is scaled differently
from the spatial coordinates.) In other words, $V$ should be such that
Eq. (38) can be written as
$$
{{d{\xi^{\prime}}^{i}}\over {dt}}(1+\epsilon h_H ) =(\omega^{-1}(\xi^{\prime},
g'))^{ij}
\left[ {{\partial H}\over {\partial \xi^j}}\right]_{\xi' ,g'} \eqno(39)
$$
Expanding to first order in $\epsilon$ and using Eq. (38), we get
$$
P^i_V ~+P^k_H {{\partial v^i}\over {\partial \xi^k}} -
v^k{{\partial P^i_H}\over {\partial \xi^k}} ~-\beta^\alpha {{\partial P^i_H}
\over {\partial g^\alpha }} +h_H P^i_H ~=0 \eqno(40)
$$
where $P^k_f=(\omega^{-1})^{kl}{{\partial f}\over {\partial \xi^{l}}}$ for
$f= H,V$. $P^i_f$ is the Hamiltonian vector field corresponding to the
phase space function $f$.
Eq. (40) is the basic expression of the invariance under RG-flow. For
many cases of interest, such as our field theory example, $v^i(\xi)$ are taken
to be linear in $\xi^i$, i.e., $v^i(\xi)= D^i_k\xi^k = (D\xi )^i$ where
$D^i_k$ are constants. We also have constant $\omega_{ij}$ with
$$
-\omega_{ji}D^{i}_{k}(\omega^{-1})^{kl}=D^{l}_{j} \eqno(41)
$$
In this case, we can multiply Eq. (40) by $\omega_{ji}$ and simplify to
obtain
$$
{{\partial}\over {\partial \xi^{j}}}\left[ V - \beta^{\alpha}{{\partial H}\over
{\partial g^{\alpha}}}H
-(D \xi)^{k} {{\partial H}\over {\partial \xi^{k}}} +~h_H ~ H \right] ~=0
\eqno(42)
$$
Thus an RG equation for H would read
$$
-{{d H}\over {d (\log \Lambda )}} - (D \xi)^{k} {{\partial H}\over
{\partial \xi^{k}}}
- \beta^{\alpha} {{\partial H}\over {\partial g^{\alpha}}} +h_H~H ~=0 \eqno(43)
$$
The term involving $(D\xi)^k$ can be written as the Poisson bracket $\{D, H\}$
where $D$ is the canonical generator of the transformation. For example,
$D= \int d^3x~ \pi (x\cdot \nabla +1+\gamma +\sigma)\varphi$ for the scalar
field theory of Eq. (1).  Eq. (43) can thus be written as
$$
-{dH \over {d(\log \L)}} +\{ D,H \}~+\beta^\alpha {\partial H \over
{\partial g^\alpha}} +h_H H ~=0\eqno(44)
$$
At this stage, one can also go to the quantum version in the standard way by
replacing Poisson brackets by commutators. The resulting equation can
also be derived from the compatibility of Schr\"odinger equation and the
RG-equation (25).
\vskip .1in

We thank R.Jackiw, P.Orland and B.Sakita for discussions.
This work was supported in part by the NSF grant  PHY 90-20495 and by the
Professional Staff Board of Higher Education of
the City University of New York under grant no. 6-63351.
\vskip .2in
\noindent{\bf References}
\vskip .2in
\item{1.} For a nice review consult R.Jackiw's article in {\sl Field Theory and
Particle Physics}, O.Eboli, M.Gomes and A.Samtoro, eds. (World Scientific,
Singapore, 1990); See also, F.Cooper and E.Mottola, Phys.Rev. {\bf D36}, 3114
(1987);
S.-Y.Pi and M.Samiullah, Phys.Rev. {\bf D36}, 3128 (1987); R.Floreanini and
R.Jackiw, Phys.Rev. {\bf D37}, 2206 (1988).
\vskip .1in
\item{2.} R.J.Perry, A.Harindranath and K.G.Wilson, Phys.Rev.Lett. {\bf 65},
2559 (1990);
R.J.Perry and  K.G.Wilson,  Nucl.Phys. {\bf  403}, 587 (1993); R.J.Perry, Ohio
preprint,
OSU-NT-93-117 (to appear in Ann. of Phys.).
\vskip .1in
\item{3.} A.H.Mueller, Nucl.Phys. {\bf B415}, 373 (1994).
\vskip .1in
\item{4.} K.G.Wilson and J.Kogut, Phys.Rep. {\bf 12},  75 (1975).
\vskip .1in
\item{5.} K.Symanzik, Nucl.Phys. {\bf B190}, 1 (1980).
\vskip .1in
\item{6.} M.L\"uscher, Nucl.Phys. {\bf B254}, 52 (1985).
\vskip .1in
\item{7.} R.P.Feynman, in Proceedings of 1987 Wangerooge Conference on {\sl
Variational Calculations in Quantum Field Theory}, pp. 28-40.
\vskip .1in
\item{8.} S.Weinberg, Nucl.Phys. {\bf B413}, 567  (1994); A.A.Abrikosov,
L.P.Gorkov and I.E.Dzyaloshinski,  {\sl Methods of Quantum Field Theory in
Statistical Physics} (Dover, 1975).
\vskip .1in
\end